\documentclass[english]{scrartcl}
\usepackage[T1]{fontenc}
\usepackage[utf8x]{inputenc}
\usepackage{xcolor}
\usepackage{pdfcolmk}
\usepackage{babel}
\usepackage{array}
\usepackage{mathrsfs}
\usepackage{url}
\usepackage{amsmath}
\usepackage{stmaryrd}
\usepackage{graphicx}
\PassOptionsToPackage{normalem}{ulem}
\usepackage{ulem}
\usepackage[unicode=true,pdfusetitle,
 bookmarks=true,bookmarksnumbered=false,bookmarksopen=false,
 breaklinks=false,pdfborder={0 0 1},backref=false,colorlinks=false]
 {hyperref}

\makeatletter

\providecommand{\tabularnewline}{\\}
\providecolor{lyxadded}{rgb}{0,0,1}
\providecolor{lyxdeleted}{rgb}{1,0,0}

\DeclareRobustCommand{\lyxdeleted}[3]{{\texorpdfstring{\color{lyxdeleted}\sout{#3}}{}}}

\newcommand{\lyxaddress}[1]{
\par {\raggedright #1
\vspace{1.4em}
\noindent\par}
}

\makeatother

\begin{document}

\title{Geometry dependence of 2-dimensional space-charge-limited currents\thanks{ $\protect\varcopyright$ 2016. This manuscript version is made available
under the CC-BY-NC-ND 4.0 license \protect\url{http://creativecommons.org/licenses/by-nc-nd/4.0/}}}

\author{Patrick De Visschere}
\maketitle

\lyxaddress{\begin{center}
Ghent University, Dept. ELIS, Liquid Crystals \& Photonics\\
Technologiepark-{}-Zwijnaarde 15, B-9052 Gent\\
Patrick.DeVisschere@UGent.be
\par\end{center}}
\begin{abstract}
The space-charge-limited current in a zero thickness planar thin film
depends on the geometry of the electrodes. We present a theory which
is to a large extent analytical and applicable to many different lay-outs.
We show that a space-charge-limited current can only be sustained
if the emitting electrode induces a singularity in the field and if
the singularity induced by the collecting electrode is not too strong.
For those lay-outs where no space-charge-limited current can be sustained
for a zero thickness film, the real thickness of the film must be
taken into account using a numerical model.
\end{abstract}

\section{Introduction}

\begin{figure}
\begin{centering}
\includegraphics[width=5cm]{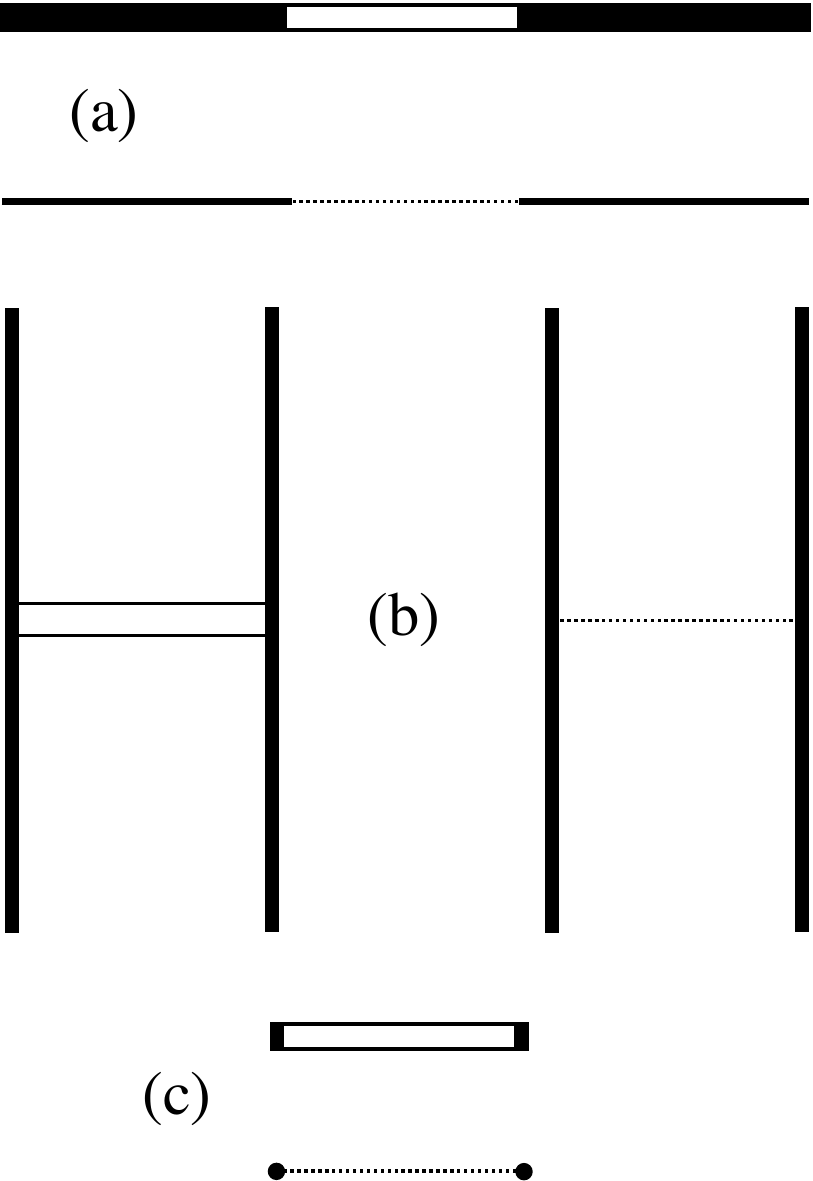}
\par\end{centering}
\caption{\label{fig:Grinberg geometries}Different 2D thin film lay-outs considered
by Grinberg e.a. \cite{Grinberg:1989fj}: (a) ``strip'' lay-out,
(b) ``plane'' lay-out and (c) ``edge'' lay-out. For each lay-out
the actual lay-out with a non-zero film thickness is shown next to
the idealized one with a zero thickness thin film and which is used
in their model.}
\end{figure}

When charge carriers are injected into an electrically poorly conducting
medium, the current is space-charge-limited and when the medium has
Ohmic conductivity, with increasing voltage, the current eventually
becomes also space-charge-limited. This phenomenon has been known
for a long time in a one-dimensional (1D) setting as described by
the Mott-Gurney equation \cite{Lampert:1970rz}
\begin{equation}
J=\frac{9}{8}\mu\epsilon\frac{V^{2}}{L^{3}}\label{eq:1}
\end{equation}

with $V$ the voltage, $J$ the current density, $L$ the width of
the insulator, $\epsilon$ it's dielectric constant and $\mu$ the
mobility of the carriers. Eq.~(\ref{eq:1}) holds in particular for
single carrier injection under perfect injection conditions, meaning
that the electric field is zero at the injecting electrode. Similar
behavior has been observed in a planar two-dimensional (2D) setting
in organic thin films \cite{Ho:2009un,Woestenborghs:2012gd,Woestenborghs:2013ep,Woestenborghs:2012ty}
and more recently in several types of monolayers \cite{Mahvash:2015qv,Dubacheva:2014ys}.
In \cite{Visschere:2015kq} we derived the following 2D version of
eq.~(\ref{eq:1}) for an infinitesimally thin layer between two semi-infinite
co-planar electrodes
\begin{equation}
K=\frac{2}{\pi}\epsilon\mu\frac{V^{2}}{L^{2}}\label{eq:2}
\end{equation}

where $K$ is the surface current density, and similar additional
results were also obtained for a photoconductor. Subsequently we discovered
a paper by Grinberg e.a. \cite{Grinberg:1989fj} where besides this
``strip'' lay-out two more lay-outs were considered: a thin film
between two parallel electrodes perpendicular to the film (``plane''
lay-out) and a thin film with small ``edge'' electrodes. These lay-outs
are shown in Fig.~\ref{fig:Grinberg geometries} together with the
idealized models used to calculate the current. Indeed only ``the
limiting case of a vanishing film thickness'' was considered and
the relevant equations were solved numerically with the aim to obtain
the prefactor $\alpha$ occurring in the general expression
\begin{equation}
K=\alpha\epsilon\mu\frac{V^{2}}{L^{2}}\label{eq:3}
\end{equation}

They found respectively $\alpha_{\mathrm{strip}}\approx0.7$, $\alpha_{\mathrm{plane}}\approx1$
and $\alpha_{\mathrm{edge}}\approx0.57$. When we applied our analytical
method to these idealized ``plane'' and ``edge'' models we found
that actually $\alpha_{\mathrm{plane}}=\alpha_{\mathrm{edge}}=0$,
meaning that in these idealized structures no space-charge-limited
(SCL) current can be sustained and to obtain a practical result the
film thickness must be taken into account.

In this paper we explore the dependence of the prefactor $\alpha$
in (\ref{eq:3}) on the lay-out systematically and analytically as
much as possible. In section~\ref{sec:Semi-infinite-coplanar-electrode}
we explain our method by deriving the value of $\alpha_{\mathrm{strip}}$
for the reference case of two semi-infinite coplanar electrodes. In
section~\ref{sec:conformally-similar} this result is extended to
other lay-outs using conformal transformations and as a result we
obtain several limits leading to the zero result for the ``plane''
lay-out. In section~\ref{sec:Non-vanishing-film-thickness} we consider
an approximate and numerical model for a thin film between planar
electrodes but with a non-zero thickness. In section~\ref{sec:Finite-electrodes}
we turn our attention to electrodes with finite width, in particular
the idealized ``edge'' electrodes. In this case a slightly different
method must be used and a single numerical integration is required
to find $\alpha$. In the last section~\ref{sec:Asymmetric-electrodes}
we consider asymmetrical lay-outs.

In their paper Grinberg e.a. refer to a paper by Geurst \cite{Geurst:1966kq}
where the exact expression $\frac{2}{\pi}$ for the prefactor occurring
in (\ref{eq:2}) for the ``strip'' lay-out was derived, as far as
we know, for the first time. This result was found by solving analytically
a boundary value problem for the square of the complex electric field.
In our method \cite{Visschere:2015kq} the problem is reduced to solving
a non-linear integral equation with a known solution, which was published
by Peters \cite{Peters:1963vi}. We will also show how these two methods
are related. A totally different approach to the problem, based on
E-Infinity theory, was published by Zmeskal e.a. \cite{Zmeskal:2007rw}.

Eq.~(\ref{eq:1})-(\ref{eq:3}) and the rest of this paper holds
for drift transport. For ballistic transport eq.~(\ref{eq:1}) must
be replaced by the (1D) \emph{Child-Langmuir }law. Some studies of
2D versions of the \emph{Child-Langmuir} law have been published for
parallel electrodes \cite{Luginsland:1996gw,Luginsland:2002ip,Umstattd:2001jh,Watrous:2001fj,Beznogov:2013yg}.
In what follows we consider the injection of positive charges from
the anode (emitter) to the cathode (collector) but the results are
obviously equally valid for negative charges.

\section{\label{sec:Semi-infinite-coplanar-electrode}Semi-infinite coplanar
electrodes}

Photoconductors are often contacted by two interdigitated electrodes
and if the fingers are much wider than the gaps in between then this
lay-out can be approximated well by two semi-infinite coplanar electrodes
as shown in Fig.~\ref{fig:Canonical-geometry-for}.
\begin{center}
\begin{figure}
\begin{centering}
\includegraphics[width=6cm]{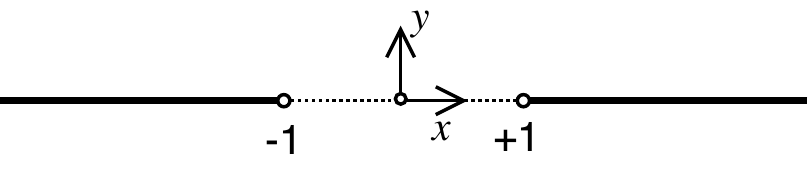}
\par\end{centering}
\caption{\label{fig:Canonical-geometry-for}Idealized ``strip'' lay-out for
2D SCL current flow. The electrodes are shown as thick lines and the
actual channel where current flows by the broken line. The small circles
have no physical meaning and are used to mark specific points only.
We use coordinates $\left(x,y\right)$ as indicated with the complex
variable $z=x+jy$.}
\end{figure}
\par\end{center}

In the calculations we will use only normalized quantities with the
channel width $L=2$ and the applied voltage $V=1$. The true surface
current density $K$ is then written as
\begin{equation}
K=2\epsilon\mu\frac{\rho}{2\epsilon}E_{x}\frac{4V^{2}}{L^{2}}\label{eq:4}
\end{equation}

where the in-plane component of the electric field $E_{x}$ and the
upper out-of-plane component $E_{y}^{+}=\frac{\rho}{2\epsilon}$,
with $\rho$ the surface charge density, are normalized by $2V/L$.
Comparing with (\ref{eq:3}) we then find the prefactor from the equation
\begin{equation}
\alpha=8E_{y}^{+}E_{x}\label{eq:5}
\end{equation}

where the field components must still satisfy Maxwell's equations.
Assuming $E_{y}^{+}$ known for all $x$, and using the Green's function,
$E_{x}$ is easily found as
\begin{equation}
E_{x}(x)=\frac{1}{\pi}\int_{-\infty}^{+\infty}\frac{E_{y}^{+}(t)}{x-t}dt\label{eq:6}
\end{equation}

where the integral is a Cauchy principal value integral. From this
equation we learn that both field components are connected by a \emph{Hilbert}-transform
over the real axis. Since the \emph{Hilbert}-transform equals it's
own inverse, except for a sign reversal, we find immediately
\begin{equation}
E_{y}^{+}(x)=\frac{1}{\pi}\int_{-1}^{+1}\frac{E_{x}(t)}{t-x}dt\label{eq:7}
\end{equation}

where we also used the boundary condition that along the electrodes
$E_{x}=0$. Substituting (\ref{eq:7}) in (\ref{eq:5}) we find that
the unknown function $\phi=E_{x}$ most be chosen in such a way that
the following expression
\begin{equation}
\alpha=\frac{8}{\pi}\phi(x)\int_{-1}^{+1}\frac{\phi(t)}{t-x}dt\label{eq:8}
\end{equation}

is a constant within the gap $-1<x<1$ and zero elsewhere. This type
of equation can be solved analytically \cite{Visschere:2015kq,Peters:1963vi}
but to obtain $\alpha$ the explicit solution is not needed (in section~\ref{sec:Finite-electrodes}
we explain how the field components can be obtained). It suffices
to integrate (\ref{eq:8}) over the gap after removing possible singularities.
This condition is necessary for reversing the order of integration
in the rhs\footnote{Formally $\int_{a}^{b}\phi_{1}(x)dx\int_{a}^{b}\frac{\phi_{2}(t)}{t-x}dt=\int_{a}^{b}\phi_{2}(t)dt\int_{a}^{b}\frac{\phi_{2}(x)}{t-x}dx$
if $\phi_{1}\in L_{p_{1}}$, $\phi_{2}\in L_{p_{2}}$ with $p_{1}^{-1}+p_{2}^{-1}\leq1$
\cite{Tricomi:1985zl}. Since $p_{2}<2$ we need $p_{1}>2$.}. In this particular case the in-plane component of the electric field
shows a singularity near $x=1$ only, whereas $E_{x}(-1)=0$ because
of the perfect injection boundary condition. Multiplying eq.~(\ref{eq:8})
with the factor $(1-x)$ and integrating we obtain
\begin{equation}
\alpha=\frac{4}{\pi}\int_{-1}^{+1}\phi(x)(1-x)dx\int_{-1}^{+1}\frac{\phi(t)}{t-x}dt\label{eq:9}
\end{equation}

Reversing the order of integration and splitting the last integral
by rewriting $(1-x)$ as $(1-t+t-x)$\lyxdeleted{Patrick De Visschere}{Tue Sep 27 09:22:11 2016}{
} then yields 
\begin{multline}
\alpha=\frac{4}{\pi}\int_{-1}^{+1}\phi(t)(1-t)dt\int_{-1}^{+1}\frac{\phi(x)}{t-x}dx\\
+\frac{4}{\pi}\int_{-1}^{+1}\phi(t)dt\int_{-1}^{+1}\phi(x)dx\label{eq:11}
\end{multline}

According to (\ref{eq:9}) the first term on the rhs equals $-\alpha$
and from the boundary condition we know that $\int_{-1}^{+1}\phi(x)dx=1$
and we find $\alpha=2/\pi$.

\section{\label{sec:conformally-similar}Lay-outs conformally similar with
2 semi-infinite coplanar electrodes}

\begin{figure}
\begin{centering}
\includegraphics[width=8.8cm]{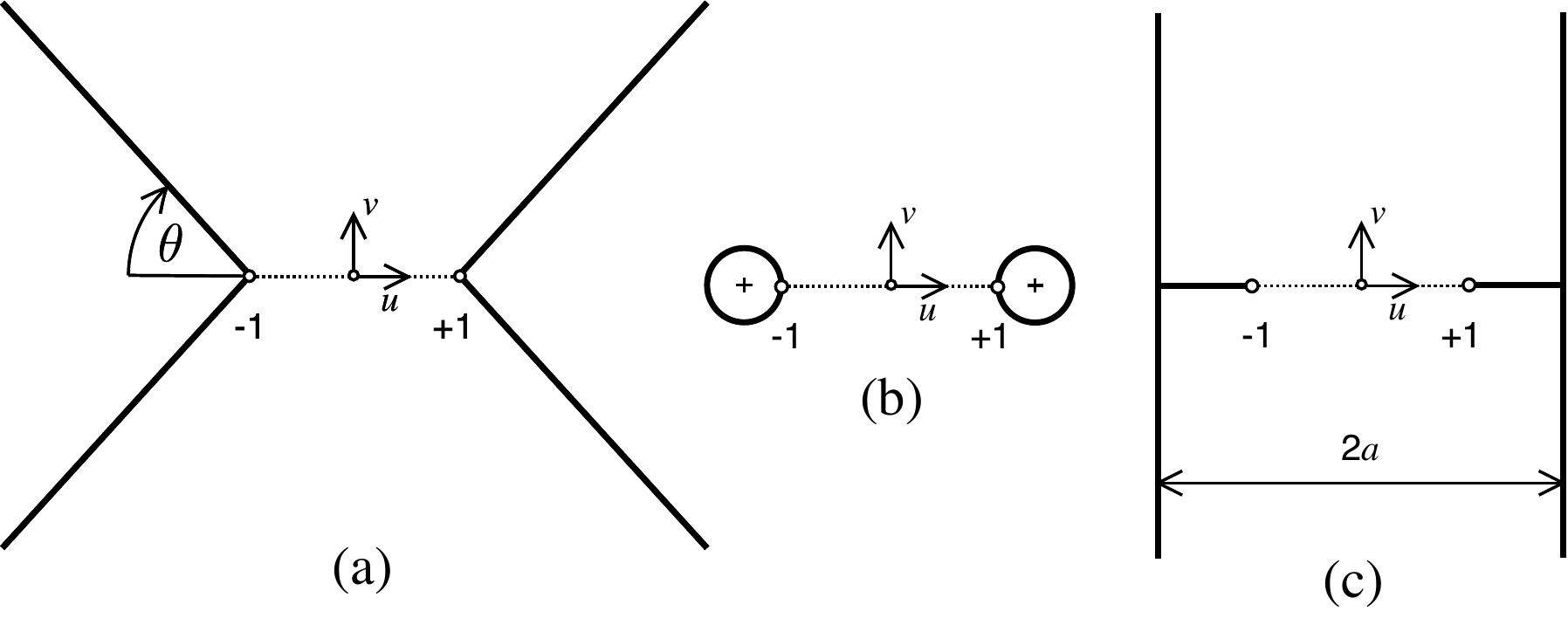}
\par\end{centering}
\caption{\label{fig:bend_electrodes}Three idealized lay-outs which are conformally
similar to the reference lay-out in Fig.~\ref{fig:Canonical-geometry-for}.
Here we use coordinates $\left(u,v\right)$ as indicated with the
complex variable $w=u+jv$.}
\end{figure}

Since we are dealing with a two-dimensional field problem, we can
use conformal transformations to obtain the field for additional lay-outs
of the electrodes \cite{Kober:1957kx,Binns:1963fj}. A straightforward
generalization of the reference structure in Fig.~\ref{fig:Canonical-geometry-for}
and which is also conformally similar is shown in Fig.~\ref{fig:bend_electrodes}a,
with $0\leq\theta\leq\pi/2$. Using $\left(u,v\right)$ coordinates
eq.~(\ref{eq:5}) is still valid after replacing the coordinates
\begin{equation}
\alpha=8E_{v}^{+}E_{u}\label{eq:12}
\end{equation}

but the relation between both field components is now more complicated.
However by transforming the lay-out of Fig.~\ref{fig:bend_electrodes}a
to the reference lay-out of Fig.~\ref{fig:Canonical-geometry-for}
by a conformal transformation we can solve the equation in the $\left(x,y\right)$-domain
of Fig.~\ref{fig:Canonical-geometry-for} instead. Indeed from the
conformality it follows that voltage drop as well as the charge density
are conserved meaning that along the real axes
\begin{equation}
E_{u}(u)du=E_{x}(x)dx
\end{equation}
\begin{equation}
E_{v}^{+}(u)du=E_{y}^{+}(x)dx
\end{equation}

and eq.~(\ref{eq:12}) becomes
\begin{equation}
\alpha\left(\frac{du}{dx}\right)^{2}=8E_{y}^{+}E_{x}
\end{equation}

where the latter must be solved in the reference lay-out of Fig.~\ref{fig:Canonical-geometry-for}
for which the field solutions (\ref{eq:6}) and (\ref{eq:7}) can
be used. This equation can still be solved with the technique of Peters
\cite{Peters:1963vi} and in particular we obtain after integration
over the channel
\begin{equation}
\alpha\int_{-1}^{+1}(1-x)\left(\frac{du}{dx}\right)^{2}dx=\frac{4}{\pi}\label{eq:16}
\end{equation}

\begin{figure}
\begin{centering}
\includegraphics[width=8cm]{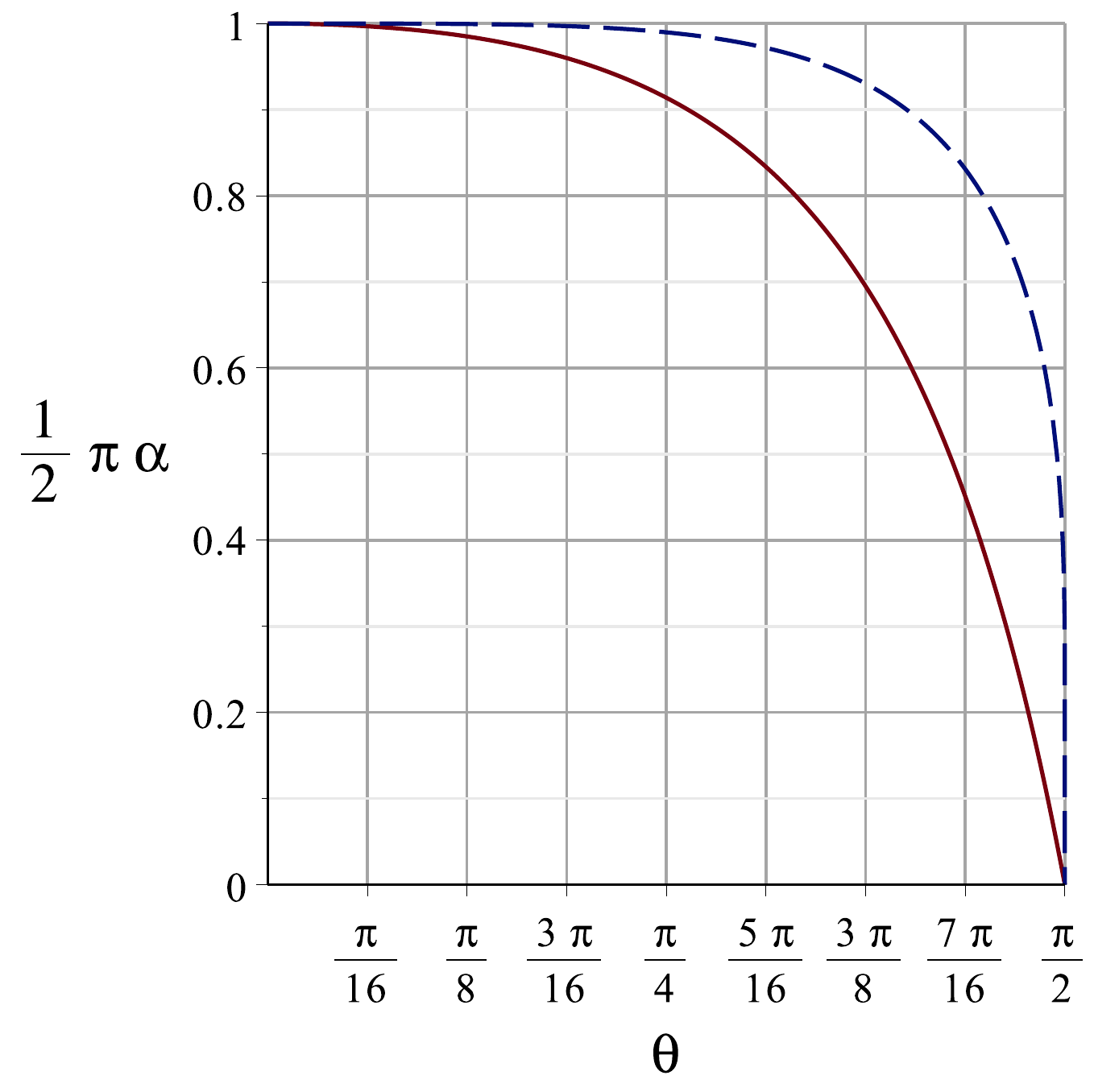}
\par\end{centering}
\caption{\label{fig:The-pre-factor-for}The prefactor for the SCL current between
two sharply bend ``wedge'' electrodes as in Fig.~\ref{fig:bend_electrodes}a
as a function of the halve opening angle $\theta$ (full line) and
a similar curve (broken line) for the lay-out shown in Fig.~\ref{fig:bend_electrodes}c
with in this case $\theta=\pi/2a$, $2a$ being the distance between
the parallel plates.}
\end{figure}

The prefactor $\alpha$ is thus easily found for every lay-out conformal
with the reference lay-out, supposed the transformation $du/dx$ for
the channel can be found. For the sharply bend (``wedge'') electrodes
in Fig.~\ref{fig:bend_electrodes}a the conformal transformation
can be found in terms of the hypergeometric function with in particular
\begin{equation}
\frac{du}{dx}=\frac{2}{\sqrt{\pi}}\frac{\Gamma\left(\frac{3}{2}-\frac{\theta}{\pi}\right)}{\Gamma\left(1-\frac{\theta}{\pi}\right)}\left(1-x^{2}\right)^{-\frac{\theta}{\pi}}
\end{equation}

where $\Gamma\left(\right)$ is the gamma function. We then obtain
$\alpha$ applying (\ref{eq:16}) 
\begin{equation}
\alpha(\theta)=\frac{1}{\sqrt{\pi}}\frac{\Gamma^{2}\left(1-\frac{\theta}{\pi}\right)}{\Gamma^{2}\left(\frac{3}{2}-\frac{\theta}{\pi}\right)}\frac{\Gamma\left(\frac{3}{2}-\frac{2\theta}{\pi}\right)}{\Gamma\left(1-\frac{2\theta}{\pi}\right)}\label{eq:symmetric_wedges}
\end{equation}

with $\alpha(0)=2/\pi$. As can be seen in Fig.~\ref{fig:The-pre-factor-for}
the value of $\alpha$ drops relatively slowly for small values of
$\theta$ and for a total opening angle of $2\theta=\pi/2$ the value
of $\alpha$ is still nearly 90\% of it's maximum value. When the
bend angle approaches zero, the value drops fast to zero. Both limits
are given by
\begin{align}
\lim_{\theta\rightarrow0}\frac{\pi}{2}\alpha & =1-\frac{1}{3}\left(\frac{12}{\pi^{2}}-1\right)\theta^{2}+O(\theta^{3})\\
\lim_{\theta\rightarrow\frac{\pi}{2}}\alpha & =\pi-2\theta+O\left(\left(\theta-\frac{\pi}{2}\right)^{2}\right)
\end{align}

For $\theta=\pi/2$ the singularity in the field due to the electrode
disappears and apparently no stable SCL current can be sustained without
this singularity. This can be understood as follows: the field near
the anode due to the space charge only, is directed towards the anode
and this field as well as the charge density itself diverge. However
the total charge induced in the anode is finite and if the anode is
smooth the field due to the charge on the anode remains finite and
therefore cannot overcome the diverging field due to the space charge.
Stationary emission of holes into the gap is only possible if the
field due to the charge on the anode also diverges. This will also
be true for any pair of electrodes with a smooth surface, like e.g.
the circular cylinders shown in Fig.~\ref{fig:bend_electrodes}b.
In this case the conformal transformation yields
\begin{equation}
\frac{du}{dx}=\frac{1}{\pi}\frac{B\sqrt{1+2r}}{\cosh^{2}\left(\frac{B}{\pi}\arcsin x\right)}\frac{1}{\sqrt{1-x^{2}}}
\end{equation}

where $B$ is a constant depending on the radius $r$ of the cylinders.
Due to the last factor the integral in the rhs of (\ref{eq:16}) diverges
and $\alpha=0$.

Another lay-out with the required singularity and with a limit leading
to the idealized ``plane'' lay-out is shown in Fig.~\ref{fig:bend_electrodes}c
for which one finds with a simple \emph{Schwartz-Christoffel} transformation
that
\begin{equation}
\frac{du}{dx}=\frac{\frac{2}{\pi}a}{\sqrt{b^{2}-x^{2}}}
\end{equation}

with $b^{-1}=\sin\frac{\pi}{2a}$. With (\ref{eq:16}) we then find
\begin{equation}
\alpha(\theta)=\frac{4}{\pi}\frac{\theta^{2}}{\sin\theta\ln\frac{1+\sin\theta}{1-\sin\theta}}
\end{equation}

with $\theta=\frac{\pi}{2a}$. In this case the drop of $\alpha$
with increasing $\theta$ is even more robust with $\alpha$ staying
above 90\% of it's limiting value as long as the combined width of
the electrode extensions is at least 25\% of the gap width ($2a\approx2.5$).
On the other hand the value of $\alpha$ drops much faster to zero
if the extension becomes much shorter. Both limits are given by
\begin{equation}
\lim_{\theta\rightarrow0}\frac{\pi}{2}\alpha=1-\frac{\theta^{4}}{45}+O(\theta^{6})
\end{equation}
\begin{equation}
\lim_{a\rightarrow1}\alpha=\frac{\pi}{2\ln\frac{4}{\pi(a-1)}}+O(a-1)
\end{equation}

We have shown that for the idealized ``plane'' lay-out, shown on
the right side in Fig.~\ref{fig:Grinberg geometries}b, no stable
SCL current can be sustained and therefore this is not an adequate
model for the practical ``plane'' lay-out shown on the left. To
calculate the SCL current for a thin film between parallel electrodes,
the thickness of the film must necessarily be taken into account and
for best results this requires a full blown numerical 2D model.

\section{\label{sec:Non-vanishing-film-thickness}Non-vanishing film thickness}

\begin{figure}
\begin{centering}
\includegraphics[width=3cm]{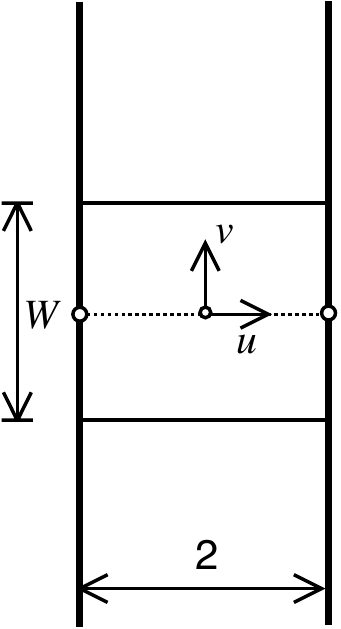}
\par\end{centering}
\caption{\label{fig:W}The practical ``plane'' lay-out: a film with thickness
$W$ between 2 parallel electrodes. }
\end{figure}

However for a film with finite thickness between parallel electrodes
(Fig.~\ref{fig:W}), an approximate value can be obtained resorting
to a one-dimensional numerical calculation only, if we neglect the
variation of the space charge density in the direction perpendicular
to the film \cite{Lau:2001jp}.

The electric field of a unit line charge placed between 2 parallel
electrodes at zero potential, and taken in the direction perpendicular
to these plates is given by \cite{Grinberg:1989fj,Kunz:1921db}
\begin{equation}
E_{u}=\frac{1}{4\epsilon}\frac{\cos\frac{\pi}{2}u'\left(\cosh\frac{\pi}{2}v'\sin\frac{\pi}{2}u-\sin\frac{\pi}{2}u'\right)}{\left(\cosh\frac{\pi}{2}v'-\sin\frac{\pi}{2}u\sin\frac{\pi}{2}u'\right)^{2}-\cos^{2}\frac{\pi}{2}u\cos^{2}\frac{\pi}{2}u'}
\end{equation}

where $\left(u,0\right)$ are the coordinates of the field point and
$\left(u',v'\right)$ those of the source point. Considering a film
$-\frac{W}{2}\leq v'\leq\frac{W}{2}$ and assuming that the charge
density is uniform in the $v'$-direction, we calculate the average
Green's function by averaging this field 
\begin{multline}
G_{W}\left(u;u'\right)=\\
\frac{1}{2W}\int_{-\frac{W}{2}}^{\frac{W}{2}}\frac{\cos\frac{\pi}{2}u'\left(\cosh\frac{\pi}{2}v'\sin\frac{\pi}{2}u-\sin\frac{\pi}{2}u'\right)}{\left(\cosh\frac{\pi}{2}v'-\sin\frac{\pi}{2}u\sin\frac{\pi}{2}u'\right)^{2}-\cos^{2}\frac{\pi}{2}u\cos^{2}\frac{\pi}{2}u'}dv'
\end{multline}

where as before we moved the factor $\frac{1}{2\epsilon}$ into the
charge density, and where the latter is defined per unit area, hence
the additional factor $W^{-1}$. This expression eventually leads
to the closed form expression
\begin{equation}
G_{W}\left(u;u'\right)=\frac{2}{\pi}\frac{1}{W}\arctan\frac{\sinh\frac{\pi}{4}W\cos\frac{\pi}{2}u'}{\sin\frac{\pi}{2}u-\cosh\frac{\pi}{4}W\sin\frac{\pi}{2}u'}
\end{equation}

where some care must be taken with the definition of the $\arctan$-function.
The (normalized) field on the horizontal centerline can then be written
as
\begin{equation}
E_{u}(u)=\int_{-1}^{+1}G_{W}(u;u')\frac{\rho(u')}{2\epsilon}du'+\frac{1}{2}
\end{equation}

with $\rho$ the charge density per unit area in the film. In this
way the film with thickness $W$ can again be treated as a film with
infinitesimally small thickness provided we use a modified Green's
function. Using $\phi=\rho/2\epsilon$ as the unknown function, the
prefactor $\alpha$ can be found by solving the equation
\begin{equation}
\alpha=8\phi(u)\left(\int_{-1}^{+1}G_{W}(u;u')\phi(u')du'+\frac{1}{2}\right)\label{eq:30}
\end{equation}

which should be a constant for $\left|u\right|\leq1$. Since the relation
between the charge density and the electric field is no longer a simple
\emph{Hilbert} transform, this equation must be solved numerically.
We have implemented two different methods for solving this equation:
a general method which becomes time-consuming for very small values
of $W$ and a series expansion method valid for $W\rightarrow0$.

In the first method eq.~(\ref{eq:30}) is discretized after scaling
the charge density and the electric field by $\sqrt{\alpha/8}$, so
that $\phi E_{u}=1$, and which can then be rewritten as
\begin{equation}
\frac{1}{\phi(u)}-\frac{1}{2}\int_{-1}^{+1}\frac{du}{\phi(u)}=\int_{-1}^{+1}G_{W}(u;u')\phi(u')du'
\end{equation}

Since at least formally $\frac{1}{\phi(-1)}=E_{u}(-1)=0$ this can
also be written as
\begin{equation}
\frac{1}{\phi(u)}=\int_{-1}^{+1}\left[G_{W}(u;u')-G_{W}(-1;u')\right]\phi(u')du'\label{eq:32}
\end{equation}

This is a non-linear \emph{Hammerstein} integral equation which we
have solved with a simple collocation scheme \cite{Kumar:1987rt}.
Introducing a partition $\left(u_{i}\right)_{1}^{M+1}$ of the interval
$\left[-1,+1\right]$ the unknown function $\phi(u)$ is approximated
by a sum $\sum_{i=1}^{M+1}a_{i}\psi_{i}$, where the basis functions
$\psi_{i}$ are associated with the nodes $u_{i}$. We use a linear
approximation (hat functions) except for the first node where we take
into account the divergent nature of the charge density and use $\psi_{1}=\sqrt{\frac{w_{1}}{1+u}}-\frac{1+u}{w_{1}}$,
$w_{1}$ being the width of the first element. A discrete set of equations
is obtained by evaluating (\ref{eq:32}) at the nodes $u_{i}$. These
non-linear equations are then solved with the \emph{Newton-Krylov}
solver nsoli.m\footnote{\url{http://www4.ncsu.edu/~ctk/newton/SOLVERS/nsoli.m}}.

In the 2nd method, we expand the charge density and the electric field
as a series in a small parameter $\epsilon$ (not to be confused with
the dielectric constant) as follows
\begin{equation}
\phi=\epsilon\left(1+\epsilon\phi_{1}+\epsilon^{2}\phi_{2}+\ldots\right)\label{eq:33}
\end{equation}
\begin{equation}
E_{u}=\frac{1}{2}+\epsilon E_{1}+\epsilon^{2}E_{2}+\ldots\label{eq:34}
\end{equation}

where $\phi_{i}$ and $E_{i}$ are unknown functions except for $\phi_{0}=1$,
and which are related by
\begin{equation}
E_{i}=\int_{-1}^{+1}G_{W}(u;u')\phi_{i-1}(u')du'
\end{equation}

Substituting (\ref{eq:33}) and (\ref{eq:34}) into (\ref{eq:30})
yields
\begin{equation}
\alpha=4\epsilon+4\left(\phi_{1}+2E_{1}\right)\epsilon^{2}+\ldots
\end{equation}

and keeping only the lowest order constant term we find the remaining
relations between these coefficients, e.g. $\phi_{1}=-2E_{1}$. In
this way the functions $\phi_{i}$ and $E_{i}$ can be calculated
sequentially. After terminating the series, $\epsilon$ is found by
applying the boundary condition $E_{u}(-1)=0$ , which is a polynomial
equation in $\epsilon$, with a single real root. Finally we find
$\alpha\approx4\epsilon$.

\begin{figure}
\begin{centering}
\includegraphics[width=8cm]{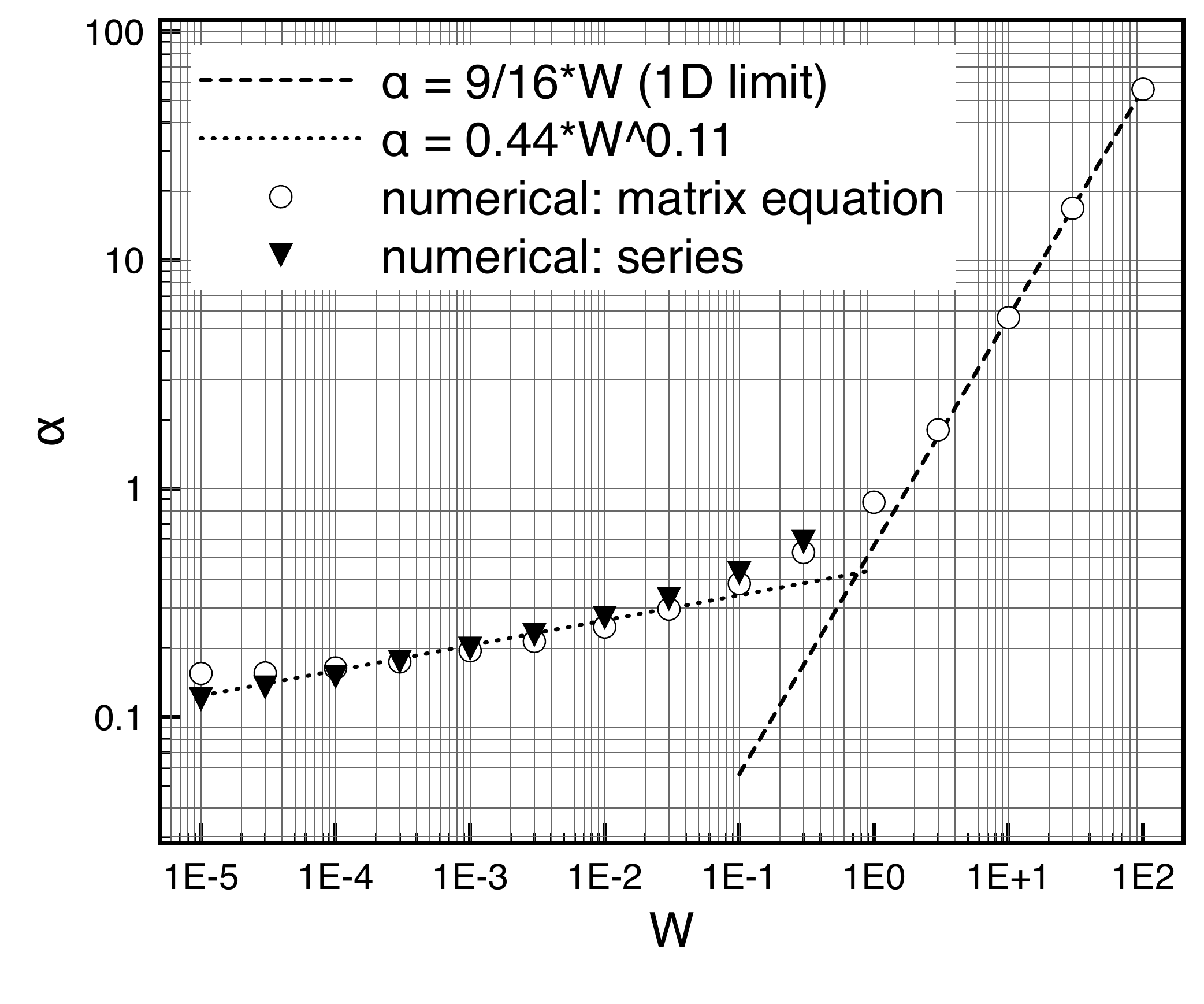}
\par\end{centering}
\caption{\label{fig:resultW}Prefactor $\alpha$ as a function of the thickness
$W$ for a film with non-zero thickness between 2 parallel plates.
The interval $\left[-1,+1\right]$ was divided into $M=110$ elements
(98 of length 0.02 and with progressive smaller elements near the
electrodes), except for the 2 smallest $W$'s where $M=210$ elements
were used. The series solution was terminated with the terms $\epsilon^{7}E_{7}$
and $\epsilon^{7}\phi_{6}$.}
\end{figure}

The results in Fig.~\ref{fig:resultW} show that the 1D-limit ($\alpha=\frac{9}{16}W$)
is approached closely if the thickness equals the width ($W\approx2$).
From consecutive approximations we've found that the series solution
approaches the solution from above and this is compatible with the
result found with the other method in the range $10^{-3}<W<1$. For
smaller $W$'s the number of elements $M=210$ is insufficient to
guarantee an accurate result with the first method. However for more
elements this method becomes prohibitively time consuming. The combined
results show that for sufficiently small $W$, $\alpha$ drops very
slowly, approximately as $\alpha\sim W^{0.11}$ and therefore that
the current density increases as $\sim W^{-0.89}$. This result should
still be handled with some caution since it is based on the assumption
of a uniform charge density and similar but also approximate calculations
for ballistic transport have shown that this is not the case \cite{Luginsland:2002ip,Watrous:2001fj}.

\section{\label{sec:Finite-electrodes}Finite width electrodes}

Returning again to a film with zero thickness, we consider in addition
electrodes with negligible dimensions. This model is used by Grinberg
e.a. \cite{Grinberg:1989fj} to simulate a thin film with ``edge''
contacts (Fig.~\ref{fig:Grinberg geometries}c). It has the advantage
that the electrodes can be treated simply as line charges. Due to
the boundary condition at the anode, the anodic line charge must be
zero and therefore the cathodic line charge must compensate the total
(in our case positive) space charge in the film. Unfortunately the
electric field of this cathodic line charge diverges and is not integrable,
meaning that, contrary to what Grinberg e.a. found, again no stable
SCL current can be sustained in this idealized structure. We arrive
at the same conclusion by considering the limit of the structure in
Fig.~\ref{fig:bend_electrodes}c when the radius of the electrodes
goes to zero. As we have seen in this case no stable space charge
can be sustained whatever the radius of the electrodes.

\begin{figure}
\begin{centering}
\includegraphics[width=6cm]{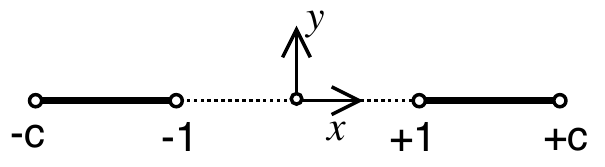}
\par\end{centering}
\caption{\label{fig:finite width electrodes}Idealized model for a thin film
between two symmetrical coplanar electrodes with finite width $c-1$.}
\end{figure}

As an alternative model for a film with small electrodes we consider
a lay-out with planar electrodes but with a finite width (Fig.~\ref{fig:finite width electrodes}),
where we expect considering the foregoing that $\lim_{c\rightarrow1}\alpha=0$.

We rewrite (\ref{eq:6}) as
\begin{equation}
E_{x}(x)=\frac{1}{\pi}\int_{-c}^{+c}\frac{E_{y}^{+}(t)}{x-t}dt\label{eq:37}
\end{equation}

since outside the electrodes ($\left|x\right|>c$) no space charge
occurs. Further we obtain
\begin{align}
8E_{y}^{+}E_{x} & =\alpha\begin{cases}
1 & \left|x\right|<1\\
0 & \left|x\right|>1
\end{cases}\label{eq:38}
\end{align}

for the same reason and since in addition on the electrodes $E_{x}=0$.
Combining (\ref{eq:37}) and (\ref{eq:38}) and setting $\phi=E_{y}^{+}$
we must solve 
\begin{equation}
\phi(x)\int_{-c}^{+c}\frac{\phi(t)}{x-t}dt=\alpha\frac{\pi}{8}\begin{cases}
1 & \left|x\right|<1\\
0 & \left|x\right|>1
\end{cases}\label{eq:39}
\end{equation}

Once more a useful relation is obtained by integrating this expression,
provided one first removes the singularities so that the order of
integration can be reversed. Since this time the charge density is
the unknown function, we must also take into account a possible singularity
for $x=-1$, therefore 
\begin{equation}
f(x)=(1-x^{2})(c^{2}-x^{2})\label{eq:40-bis}
\end{equation}
 and we then obtain
\begin{equation}
\alpha=\frac{6}{\pi}\frac{q_{1}q_{2}}{c^{2}-\frac{1}{5}}=\frac{6}{\pi}\frac{\gamma q_{1}^{2}}{c^{2}-\frac{1}{5}}\label{eq:40}
\end{equation}

where $q_{1}$ and $q_{2}$ are the first and second order (dipole
and quadrupole) moments of $\phi$ and we have taken into account
that the zeroth order moment $q_{0}=0$. We also introduced $\gamma=q_{2}/q_{1}$.
Since these moments are not known beforehand, this time (\ref{eq:40})
is not sufficient to determine $\alpha$. The complete solution can
be obtained if besides multiplying (\ref{eq:39}) with $f(x)$ we
also introduce a 2nd \emph{Hilbert} transform as follows
\begin{equation}
\int_{-c}^{+c}\frac{\phi(s)f(s)}{x-s}ds\int_{-c}^{+c}\frac{\phi(t)}{s-t}dt=\alpha\frac{\pi}{8}\int_{-1}^{+1}\frac{f(s)}{x-s}ds\label{eq:41}
\end{equation}

where $s$ is used as an additional real integration variable. The
order of the two Cauchy principal value integrals in the lhs can be
reversed by using the \emph{Poincaré-Hardy-Bertrand} theorem \cite{Tricomi:1985qf}
and after some calculations (\ref{eq:41}) can be reduced to
\begin{equation}
E_{x}^{2}-\left(E_{y}^{+}\right)^{2}=\frac{1}{f(x)}\frac{1}{\pi^{2}}\left[q_{1}^{2}-\alpha\frac{\pi}{4}\int_{-1}^{+1}\frac{f(s)}{s-x}ds\right]
\end{equation}

Using (\ref{eq:40}) and expanding the remaining integral we find
\begin{multline}
E_{x}^{2}-\left(E_{y}^{+}\right)^{2}=\\
\frac{\alpha}{2\pi}\left[\frac{x\left(c^{2}-\frac{1}{3}\right)+\frac{1}{3\gamma}\left(c^{2}-\frac{1}{5}\right)+x\left(1-x^{2}\right)}{(1-x^{2})(c^{2}-x^{2})}-\frac{1}{2}\ln\left|\frac{1-x}{1+x}\right|\right]\label{eq:43}
\end{multline}

Since $E_{x}\rightarrow0$ on both sides of $x=-1$, the rhs of (\ref{eq:43})
must remain negative when $x$ passes through the point -1 and therefore
in the rational term between brackets the pole $x=-1$ should be compensated
by a corresponding zero, meaning that
\begin{equation}
\gamma=\frac{1}{3}\frac{c^{2}-\frac{1}{5}}{c^{2}-\frac{1}{3}}
\end{equation}

and with (\ref{eq:40})
\begin{equation}
\alpha=\frac{2}{\pi}\frac{q_{1}^{2}}{c^{2}-\frac{1}{3}}\label{eq:46-bis}
\end{equation}

and finally
\begin{equation}
\left(E_{y}^{+}\right)^{2}-E_{x}^{2}=\frac{\alpha}{2\pi}\left[\frac{1}{2}\ln\left|\frac{1-x}{1+x}\right|+\frac{x^{2}-x-\left(c^{2}-\frac{1}{3}\right)}{\left(c^{2}-x^{2}\right)\left(1-x\right)}\right]\label{eq:46}
\end{equation}

It's now clear that near $x=-1$ the charge density remains integrable
and therefore we could have omitted in (\ref{eq:40-bis}) the factor
$(x+1)$ in $f(x)$. In that case $q_{2}$ and $\gamma$ would not
have occurred and instead of (\ref{eq:40}) and (\ref{eq:43}) we
would have found (\ref{eq:46-bis}) and (\ref{eq:46}) immediately.

\begin{figure}
\begin{centering}
\includegraphics[width=8cm]{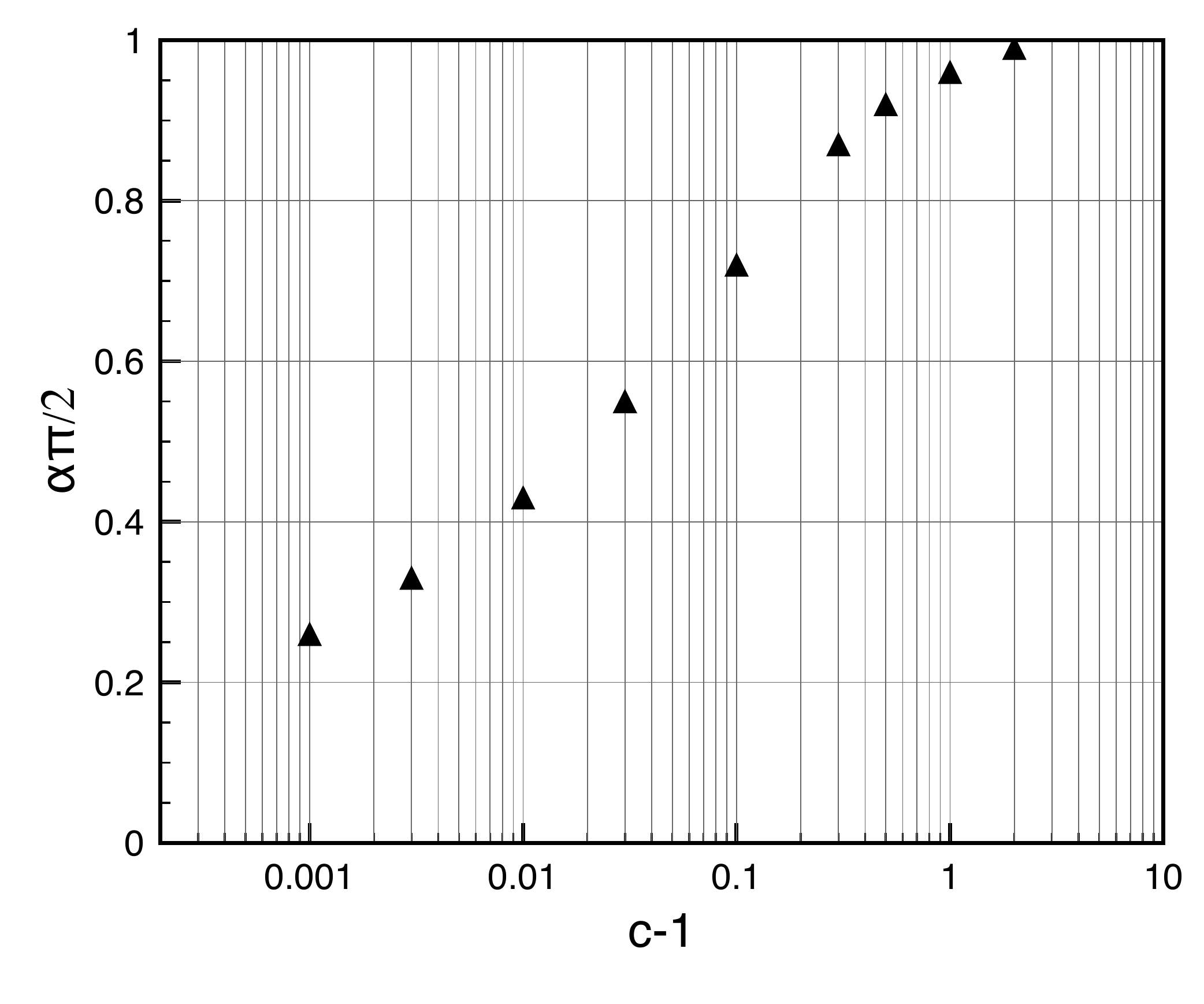}
\par\end{centering}
\caption{\label{fig:alpha versus c-1}Relative prefactor $\frac{\pi}{2}\alpha$
for the SCL current for an infinitesimally thin film between two symmetrical
electrodes with finite width $c-1$.}
\end{figure}

Combining (\ref{eq:46}) with (\ref{eq:38}) $E_{x}$ and $E_{y}^{+}$
can be calculated in the plane $y=0$ up to a scaling factor $\sim\sqrt{\alpha}\sim\left|q_{1}\right|$
which can be determined numerically by the condition that $\int_{-1}^{+1}E_{x}dx=1$.
As shown in Fig.~\ref{fig:alpha versus c-1} the value of $\alpha$
depends rather weakly on the width of the electrodes. For an electrode
width equal to the width of the thin film ($c=3)$ the SCL current
very nearly reaches it's maximum value and this value reduces to 50\%
for electrodes with a width of approximately 1\% of the width of the
thin film ($c=1.02$). 

Knowing $E_{x}$ or $E_{y}^{+}$ on the real axis ($y=0$) is sufficient
to determine the field in the rest of the plane. If we introduce the
complex electric field $\mathcal{E}(z)=E_{y}(z)+jE_{x}(z)$, then
just above and below the real axis $\mathcal{E}^{\pm}=\pm E_{y}^{+}+jE_{x}$
and after squaring $\left(\mathcal{E}^{\pm}\right)^{2}=\left(E_{y}^{+}\right)^{2}-E_{x}^{2}\pm2jE_{x}E_{y}^{+}$,
which can be obtained by combining Eqs. (\ref{eq:46}) with (\ref{eq:38}).
As explained by Peters \cite{Peters:1963vi} this expression is then
readily extended to the whole $z$-plane
\begin{equation}
\mathcal{E}^{2}=\frac{\alpha}{2\pi}\left[\frac{1}{2}\ln\frac{z-1}{z+1}+\frac{z^{2}-z-\left(c^{2}-\frac{1}{3}\right)}{\left(z^{2}-c^{2}\right)\left(z-1\right)}\right]\label{eq:47}
\end{equation}

With some hindsight we can obtain the same result much faster by applying
the method used by Geurst \cite{Geurst:1966kq} for the infinite ``strip''
lay-out ($c=\infty$). Since $\mathcal{E}^{2}$ is analytic outside
the strip $\left|x\right|\leq c,\,y=0$ and according to (\ref{eq:38})
$\Im\left(\mathscr{E}^{2}\right)$ is a known constant on this segment,
the square of the field must be of the form
\begin{equation}
\mathcal{E}^{2}=\frac{\alpha}{2\pi}\left[\frac{1}{2}\ln\frac{z-1}{z+1}+\frac{a_{2}z^{2}+a_{1}z+a_{0}}{\left(z^{2}-c^{2}\right)\left(z-1\right)}\right]\label{eq:49}
\end{equation}

where the first part solves the non-homogeneous problem and the rational
function is the most general solution of the homogeneous problem.
This rational function must tend to zero at infinity and can only
contain poles at the extremities of the electrodes, but we have immediately
taken into account that no pole occurs for $z=-1$. The unknown coefficients
$a_{i}$ and $\alpha$ can be found by considering the limit for $z\rightarrow\infty$.
With $q_{1}$ the (normalized) dipole moment one finds readily that
$\lim_{z\rightarrow\infty}\mathcal{E}^{2}=-\frac{q_{1}^{2}}{\pi^{2}z^{4}}$
and by expanding the expression between the brackets in (\ref{eq:49})
and equating corresponding terms one finds $a_{2}=1$, $a_{1}=-1$,
$a_{0}=-\left(c^{2}-\frac{1}{3}\right)$, in agreement with (\ref{eq:47}),
as well as (\ref{eq:46-bis}).

\section{\label{sec:Asymmetric-electrodes}Asymmetric electrodes}

From the foregoing we have learned that of the 3 lay-outs shown in
Fig.~\ref{fig:Grinberg geometries} only the ``strip'' lay-out
can sustain a stable SCL current. However the reasons for the absence
of a stable SCL current between (idealized) ``plane'' and ``edge''
electrodes are totally different: the ``plane'' lay-out fails because
of a missing singularity in the field of the emitting electrode, whereas
the ``edge'' lay-out fails because the singularity at the collecting
electrode is too strong. Obviously higher SCL currents can be obtained
with asymmetric lay-outs where the singularity of the emitting electrode
is as strong as possible, but where the collecting electrode is as
large as possible. This opens up 3 additional asymmetric lay-outs:
``strip/plane'', ``edge/strip'' and ``edge/plane''.

The result (\ref{eq:symmetric_wedges}) obtained for the symmetric
``wedge'' electrodes shown in Fig.~\ref{fig:bend_electrodes}a
can readily be extended to an asymmetric lay-out where the halve opening
angles of the emitting and collecting wedges are different $\theta_{a}\neq\theta_{c}$,
namely 
\begin{equation}
\alpha(a,c)=\frac{1}{\pi}\frac{\Gamma^{2}(1-a)\Gamma^{2}(1-c)\Gamma(3-2a-2c)}{\Gamma^{2}(2-a-c)\Gamma(1-2a)\Gamma(2-2c)}
\end{equation}

with $a=\theta_{a}/\pi$ and $c=\theta_{c}/\pi$. Stable emission
requires $a<\frac{1}{2}$ and besides by $\alpha(\frac{1}{2},c)=0$,
the surface is bounded by the curves
\begin{align}
\alpha(0,c) & =\frac{2}{\pi}\frac{1}{1-c}\\
\alpha(a,0) & =\frac{2}{\pi}\frac{1-2a}{1-a}\\
\alpha(a,1-a) & =\frac{2}{\tan a\pi}
\end{align}

\begin{figure}
\begin{centering}
\includegraphics[width=7cm]{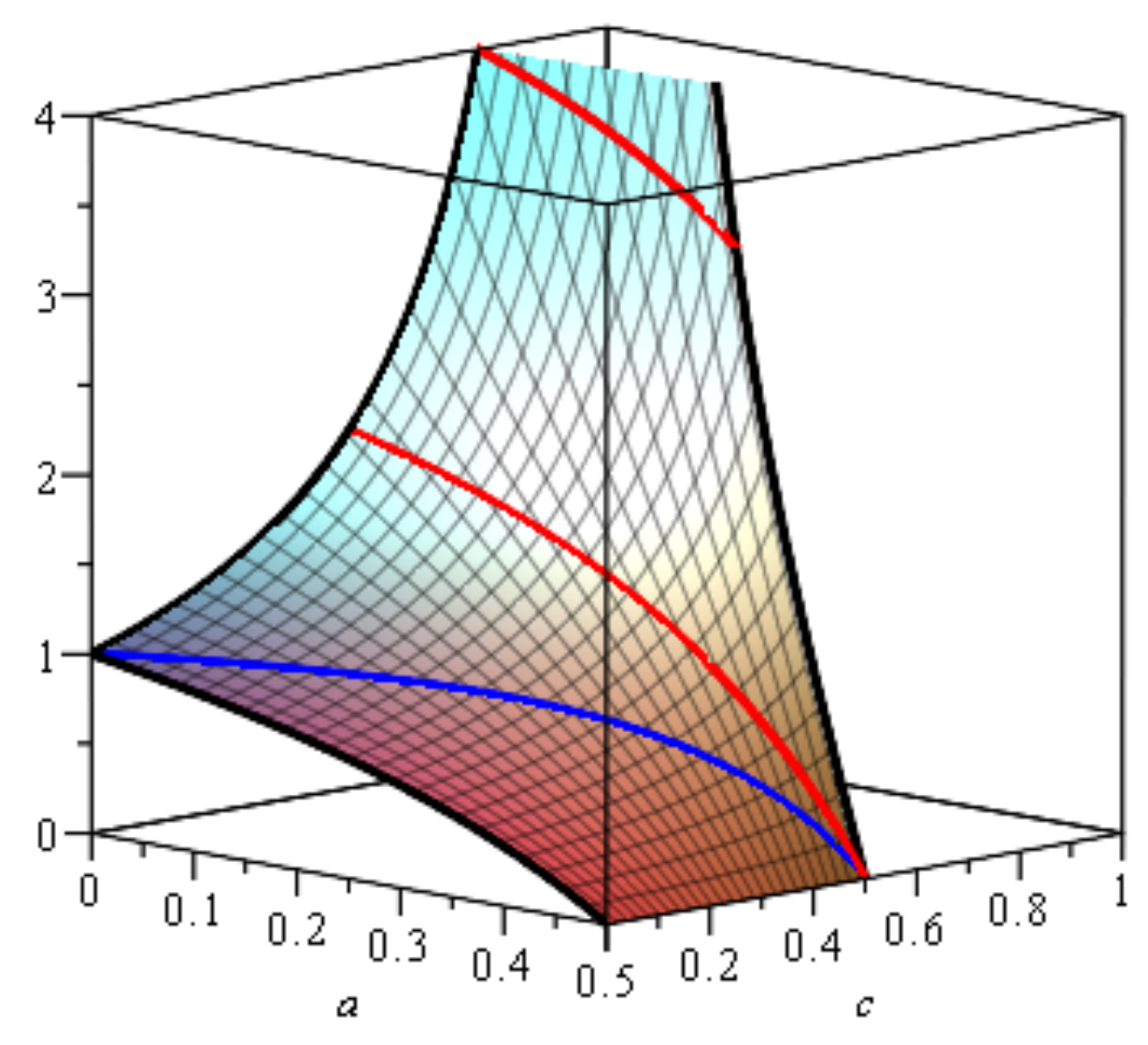}
\par\end{centering}
\caption{\label{fig:asym-wedge}Relative prefactor $\frac{\pi}{2}\alpha$ for
the SCL current between 2 ``wedge'' shaped electrodes with halve
opening angles $a\pi$ for the (emitting) anode and $c\pi$ for the
(collecting) cathode. Besides the edges of the surface ($a=0$, $c=0$
and $a+c=1$) additional curves are shown from bottom to top, in blue
for the symmetric layout $c=a$ and in red for the plane collector
($c=\frac{1}{2}$) and for a concave collector with $c=\frac{3}{4}$.
The surface has been cut at a height of 4 and tends to infinity for
$\left(a,c\right)\rightarrow\left(0,1\right)$.}
\end{figure}

As shown in Fig.~\ref{fig:asym-wedge} the surface $\alpha(a,c)$
decreases with increasing $a$ and increases with increasing $c$
confirming that for maximizing the SCL current the singularity of
the emitter should be maximal and that of the collector minimal. For
the ``strip/plane'' lay-out $\alpha_{\mathrm{strip/plane}}=\frac{4}{\pi}$
and compared with the symmetric ``strip'' electrodes, the SCL current
has doubled in value. The current can be increased further by allowing
a concave collecting electrode, e.g. for $c=\frac{3}{4}$ the current
doubles once more.

Similarly (\ref{eq:46}) valid for 2 symmetric and finite ``strip''
electrodes can be extended to the asymmetric case with the result
\begin{multline}
E_{x}^{2}-\left(E_{y}^{+}\right)^{2}=\\
\frac{\alpha}{2\pi}\left[\frac{-x^{2}+(1+c-a)x+ac-\frac{1}{3}+a-c}{(1-x)(c-x)(a+x)}-\frac{1}{2}\ln\left|\frac{1-x}{1+x}\right|\right]\label{eq:51}
\end{multline}

where the width of the emitting anode equals $a-1$ and that of the
collecting cathode $c-1$. From this equation and (\ref{eq:38}) the
fields $E_{x}$ and $E_{y}^{+}$ can be obtained again up to a scale
factor $\sim\sqrt{\alpha}$ which can then be obtained by the condition
$\int_{-1}^{1}E_{x}dx=1$. For the ``edge/strip'' lay-out with $a=1$
and $c=\infty$ we found $\alpha_{\mathrm{edge/strip}}=1.94\frac{2}{\pi}$.
By increasing the singularity of the emitting electrode from that
of an infinite ``strip'' to that of a line electrode the SCL current
has approximately doubled. We note that the specific variation of
the field components in eq. (\ref{eq:51}), has been confirmed by
a more realistic numerical model for symmetric infinite ``strip''
electrodes \cite{Visschere:2015kq}.

The final ``edge/plane'' layout can be reduced to the previous case
by a conformal transformation similar to what was done in section~\ref{sec:conformally-similar}
and yields $\alpha_{\mathrm{edge/plane}}=3.04\frac{2}{\pi}$. Again
the current can be further increased by letting the collecting electrode
become concave (see Table.~\ref{tab:Overview-of-the} where an overview
is given of selected lay-outs).

\begin{table}
\begin{centering}
\begin{tabular}{|>{\raggedleft}m{1.5cm}|>{\centering}m{1cm}|>{\centering}m{1.5cm}|>{\centering}m{1cm}|>{\centering}m{1cm}|}
\cline{2-5} 
\multicolumn{1}{>{\raggedleft}m{1.5cm}|}{} & \includegraphics{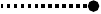} & \includegraphics[width=1.5cm]{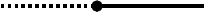} & \includegraphics{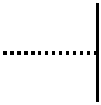} & \includegraphics{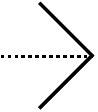}\tabularnewline
\hline 
\includegraphics{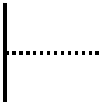} & 0 & 0 & 0 & 0\tabularnewline
\hline 
\includegraphics[width=1.5cm]{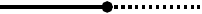} & 0 & 1 & 2 & 4\tabularnewline
\hline 
\includegraphics{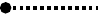} & 0 & 1.94 & 3.04 & 5.12\tabularnewline
\hline 
\end{tabular}
\par\end{centering}
\caption{\label{tab:Overview-of-the}The prefactor $\frac{\pi}{2}\alpha$ for
symmetrical and asymmetrical electrode lay-outs. The emitters on the
left are ordered with increasing singularity from top to bottom, the
collectors on the top with decreasing singularity from left to right.
The dotted line represents the insulating film carrying the space
charge.}
\end{table}

\section{Conclusions}

We have presented a method for calculating the space-charge limited
current in a 2D planar film with zero thickness between two initially
coplanar electrodes. Assuming a uniform dielectric constant the current
density is of the form (\ref{eq:2}) and the problem reduces to the
calculation of the prefactor $\alpha$. In contrast with the well-known
1D Mott-Gurney law, the value of $\alpha$ depends on the lay–out
of the electrodes and can take any value between 0 and $\infty$.
The method gives analytical expressions for both field components
in the plane of the film. Using conformal transformations the method
can be extended to non planar lay-outs of the electrodes.

We found that a stable SCL current can only be sustained if, besides
the space charge, the injecting electrode also induces a singularity
in the field and the current increases with the strength of this singularity.
In particular no stable SCL current can be sustained in a thin film
with zero thickness placed between two parallel electrodes and we
analyzed several limits leading to the zero current for this idealized
``plane'' lay-out. These results contradict previous published findings,
but since these where obtained by solving the relevant equation numerically
they where doomed to fail. As a second condition we noted that the
singularity in the field induced by the collecting electrode should
be not too strong in order for the field to remain integrable. It
follows that also between idealized ``edge'' electrodes no SCL current
can be sustained. Since the requirements for the two electrodes are
conflicting, higher SCL currents can be obtained between asymmetrical
electrodes and for a convex collecting electrode the maximum is obtained
with an ``edge'' emitter and a ``plane'' collector, the current
being 3.04 times the current between semi-infinite ``strip'' electrodes.
The current can be increased indefinitely by allowing a concave collector.

There exist some experimental evidence for the voltage and length
dependence in (\ref{eq:3}): see \cite{Visschere:2015kq} for experiments
with an organic photoconductor and \cite{Mahvash:2015qv} for experiments
on a hexagonal-BN monolayer. Unfortunately since in these experiments
the mobility usually is not known it's not possible to extract a value
for the prefactor. However it should be possible to compare different
electrode geometries using the same layer without knowing the mobility.
In particular the large polarity dependence for well chosen asymmetrical
electrodes from Table~\ref{tab:Overview-of-the} should be amenable
to experimental verification.

Besides the electrode lay-outs considered the current can be calculated
along the same lines for any lay-out which is conformally similar
to one of the lay-outs considered. A further example which comes to
mind is a periodic array of finite width electrodes. It is also likely,
although we only illustrated it for a single example, that the shortcut
introduced by Geurst \cite{Geurst:1966kq} for calculating the field
components can be applied in all those cases.

For a practical ``plane'' lay-out a non-zero thickness of the film
must necessarily be considered. We presented two approximate numerical
models to calculate the current in this case and this revealed that
the current drops relatively slowly with decreasing thickness as $\left(\frac{W}{L}\right)^{0.11}$.
However since we averaged the space charge density in the direction
perpendicular to the film, these results have to be confirmed by a
2D numerical model. This also holds for the other zero cases in Table~\ref{tab:Overview-of-the},
in particular for the practical ``edge'' lay-out. These might be
challenging problems to solve numerically if $W\ll L$ and it remains
an open question whether for those cases the limits for $W\rightarrow0$
might be found analytically or semi-numerically.

\section*{Acknowledgment}

I am grateful to Kristiaan Neyts for the stimulating and clarifying
discussions on the subject of this paper.

\bibliographystyle{elsarticle-num}

\end{document}